\documentclass[aps,prc,twocolumn,showpacs,superscriptaddress,floatfix]{revtex4}       
\usepackage{graphicx}
\newcommand{\beq}{\begin{equation}}
\newcommand{\eeq}{\end{equation}}
\begin{document}
\title{Complete fusion of $^9$Be with spherical targets}
\author{Henning Esbensen}
\affiliation{Physics Division, Argonne National Laboratory, Argonne, Illinois 60439}
\date{\today}
\def\o16{$^{16}$O}
\def\pb208{$^{208}$Pb}

\begin{abstract}
The complete fusion of $^9$Be with $^{144}$Sm and $^{208}$Pb targets is 
calculated in the coupled-channels approach. The calculations include 
couplings between the 3/2$^-$, 5/2$^-$, and 7/2$^-$ states in the
$K=3/2$ ground state rotational band of $^9$Be. 
It is shown that the $B(E2)$ values for the excitation of these states are 
accurately described in the rotor model.
The interaction of the strongly deformed $^9$Be nucleus with a spherical 
target is calculated using the double-folding technique and the effective 
M3Y interaction, which is supplemented with a repulsive term that is 
adjusted to optimize the fit to the data for the $^{144}$Sm target.
The complete fusion is described by in-going-wave boundary conditions.
The decay of the unbound excited states in $^9$Be is considered explicitly 
in the calculations by using complex excitation energies.
The model gives an excellent account of the complete fusion (CF) data for 
$^9$Be+$^{144}$Sm, and the cross sections for the decay of the excited states 
are in surprisingly good agreement with the incomplete fusion (ICF) data.
Similar calculations for $^9$Be+$^{208}$Pb explain the total fusion data at 
high energies but fail to explain the CF data, which are suppressed by 20\%, 
and the calculated cross section for the decay of excited states is a factor
of three smaller than the ICF data at high energies.
Possible reasons for these discrepancies are discussed.
\end{abstract}
\maketitle

\section{Introduction}

The influence of breakup on the complete and incomplete fusion
of weakly bound nuclei with stable targets is currently being
studied at many experimental facilities around the world. 
Experiments with unstable nuclei are particularly challenging
because of weak beam currents and poor statistics. 
Fortunately, there are several light elements that are both 
stable and weakly bound and they provide the opportunity to 
study the influence of breakup on fusion with good statistics.

A good example of a stable and weakly bound nucleus is $^9$Be 
and its fusion has been measured with $^{208}$Pb and $^{144}$Sm targets
\cite{anu99cf,anu04pb,anu06sm}. The simplest view of $^9$Be is a 
strongly deformed three-body system consisting of two $\alpha$ particles 
held together by a weakly bound neutron. 
The Q-value for the $\alpha+\alpha+n$ three-body breakup is -1.574 MeV. 
In both experiments it was possible to separate the complete from 
the incomplete fusion.

One of the most dominant features in coupled-channels calculations of the 
fusion of the strongly deformed $^9$Be nucleus with a stable target is the 
excitation of states in the $K$=3/2 ground state rotational band of $^9$Be.
The large quadrupole deformation of $^9$Be (derived from the measured
quadrupole moment of the ground state) implies that conventional 
calculations that are based on a deformed Woods-Saxon potential
may become unrealistic, for example, if the curvature corrections 
to the ion-ion potential \cite{curve}, which are due to the deformation 
of the reacting nuclei, are ignored. 
These problems are overcome in the following by using 
the double-folding technique \cite{misi1,misi2} 
to calculate the Coulomb plus nuclear interaction between the 
deformed $^9$Be nucleus and a spherical target.

Another interesting feature of $^9$Be in the non-zero spin of the 
ground state. This feature was pointed out in Ref. \cite{anu06sm},
and it was recommended that the spin of the $3/2^-$ ground state and the 
$5/2^-$ and $7/2^-$ excited states of $^9$Be should be treated explicitly 
in the calculations. 
In particular, the fusion cross section should be calculated for each 
of the initial magnetic quantum numbers, $m$ = $\pm 1/2$ and $\pm 3/2$, 
of the $3/2^-$ ground state of $^9$Be, and the average cross section 
should be compared to measurements. In the rotor model one can easily 
calculate the necessary matrix elements from the  multipole 
expansion of the total interaction between projectile and target.

The structure and the parametrization of the one-body density of $^9$Be 
is discussed in the next section. The calculation of the double-folding
interaction between $^9$Be and a spherical target is presented in Sect. III.
A model for calculating the complete fusion cross section
is presented in Sect. IV. The results of coupled-channels calculations
of the fusion of $^9$Be+$^{144}$Sm and $^9$Be+$^{208}$Pb are presented
in Sect. V, and Sect. VI contains the conclusions.

\section{Structure of $^9$Be}

The nucleus $^9$Be behaves like an almost perfect rotor with 
respect to quadrupole excitations of the $K=3/2$ ground state 
rotational band with spins $I^\pi$ = $3/2^-$, $5/2^-$, and $7/2^-$. 
This can be seen by comparing the measured, reduced transition
probabilities for quadrupole transitions to the results obtained
from the expression, Eq. (4.68a) of Ref. \cite{BMII},
\beq 
B(E2,KI\rightarrow KI') = \frac{5Q_0^2e^2}{16\pi} \ 
\langle IK \ 20|I'K\rangle^2,
\label{BE2}
\eeq 
which applies to a perfect rotor.
Here $Q_0$ is the intrinsic quadrupole moment. The measured
quadrupole moment of the $3/2^-$ ground state of $^9$Be is 
5.29(4) fm$^2$ \cite{ENDSF} which translates into the 
intrinsic quadrupole moment $Q_0$ = 26.45(20) fm$^2$, according 
to Eq. (4-69) of Ref. \cite{BMII}.
Inserting this value into Eq. (\ref{BE2}) one obtains the reduced 
transition probabilities that are shown in the last column of Table I. 
They are in good agreement with the measured values shown 
in the 4th column of Table I. The sum of the B-values is shown in the
last row of Table I. The sum is $5Q_0^2e^2/(16\pi)$ in the rotor model
and differs from the experimental value by about 5\%.

A great advantage of the rotor model is that it can be applied to 
calculate the transition matrix elements in cases where they have 
not been measured, for example, for the $5/2^-$ to $7/2^-$ transition, 
and the quadrupole moments of the excited states. One can also calculate
matrix elements of the total interaction between $^9$Be and a spherical
target from a multipole expansion of this interaction as discussed
in Sect. III. 

The excitation energies and widths of the three states are shown in
the 2nd and 3rd columns. It is noted that the width of the 7/2$^-$
state is very large which implies that the state, if excited, may
decay by particle emission before the fusion with the target takes 
place. This possibility is investigated in Sect. IV and V.

\begin{table}
\caption{The measured reduced transition probabilities
$B(E2,3/2^-\rightarrow I^-)$ for exciting the $K$=3/2 
ground state rotational band of $^9$Be \cite{ENDSF}
are compared to predictions of the rotor model. 
Also shown are the excitation energies $E_I$ and decay 
widths $\Gamma_I$ \cite{Tilley}.}
\begin{tabular} {|c|c|c|c|c|}
\colrule
 & & & experiment & rotor model \\
 spin   & $E_I$ & $\Gamma_I$ & $B(E2)$       & $B(E2)$  \\ 
$I^\pi$   & (MeV) &  (keV)   & (e$^2$fm$^4$) & (e$^2$fm$^4$) \\
\colrule
3/2$^-$ &   0    & 0    & 14.0(2) & 14 \\ 
5/2$^-$ & 2.429  & 0.78 & 40.7(6) & 36 \\ 
7/2$^-$ & 6.38   & 1210 & 18.9(3) & 20 \\ 
\colrule
sum     &        &      & 73.6(11) & 70 \\ 
\colrule
\end{tabular}
\end{table}

There are other low-lying states in $^9$Be 
but they are not expected to
play any significant role in heavy-ion collisions. Thus the spin-orbit 
partners of the ground state rotational band, i.~e., the $1/2^-$, $3/2^-$, 
and $5/2^-$ states, will be ignored because spin-excitations are weak.
The excitation of positive parity states, starting with the 
lowest $1/2^+$ state, is also weak and will be ignored.

\subsection{Density parametrization}

The densities of the deformed projectile and spherical target nuclei 
will be parametrized by the expression
\beq
\rho(r,\theta') = C \frac{1+\cosh(R/a)}{\cosh(r/a)+\cosh(R/a)},
\label{rhod} 
\eeq
where $C$ is a normalization constant, $R$ is the radius, and $a$
is the diffuseness. The radius of the deformed, axially symmetric 
projectile depends on the direction with respect to the symmetry axis.
It is parametrized as 
\beq
R(\theta') = R_0\bigl[1+ \beta_2 Y_{20}(\theta')\bigr],
\label{shaperad}
\eeq 
where $\theta'$ is the angle between the position vector ${\bf r}$ 
and the direction ${\bf e}$ of the symmetry axis.

The advantages of the parametrization (\ref{rhod}) are that it 
is similar to a Fermi function at larger values of $r$ and it 
is well behaved as a function of $\theta'$ for $r\rightarrow 0$,
where it approaches an orientation independent constant.
Another advantage of Eq. (\ref{rhod}) is the analytic properties 
it has for spherical nuclei \cite{Esb07}. For example, the Fourier 
transform is an analytic function, and the expression for the 
root-mean-square (RMS) radius,
\beq
\langle r^2 \rangle = \frac{3}{5} \bigl( R^2 + \frac{7}{3}
(a\pi)^2\bigr),
\eeq
is an exact relation (see the appendix of Ref. \cite{Esb07}.)
These features will be utilized for the spherical target nuclei,
$^{144}$Sm and $^{208}$Pb.  The parameters that have been chosen 
are shown in Table II. They were adjusted so that the measured
RMS charge radii \cite{vries} were reproduced. The parameters 
for the neutron $(\nu)$ densities were assumed to be the same as 
for protons ($\pi$), except in the case of $^{208}$Pb, where a 
slightly larger radius is used to accommodate for the neutron 
skin of this nucleus. The adopted skin thickness: 
$\delta_{np} = <r^2>_n^{1/2}-<r^2>_p^{1/2}$ $\approx$ 0.16(6) fm 
was chosen because it falls in the midst of values predicted by 
Skyrme Hartree-Fock (HF) calculations \cite{yoshida}. Moreover,
it is also consistent with the skin thickness
$\delta_{np}$ = 0.16 $\pm (0.02)_{\rm stat} \pm (0.04)_{\rm sys}$ fm
that has been extracted from antiprotonic $^{208}$Pb atoms
\cite{Klos}. The parameters for $^9$Be are determined below. 

\begin{table}
\caption{Density parameters for $^{9}$Be, $^{144}$Sm, and $^{208}$Pb.
The measured RMS charge radii \cite{vries} shown in the last column 
are reproduced, and so is the intrinsic quadrupole moment 
of $^9$Be, $Q_0$ = 26.45(20) fm$^2$ \cite{ENDSF}.
The last line shows the adopted neutron ($\nu$) density parameters
for $^{208}$Pb; the `RMS-exp' radius is estimated from 
Skyrme HF calculations \cite{yoshida}.}
\begin{tabular} {|c|c|c|c|c|c|}
\colrule
Nucleus  & R (fm) & a (fm)& $\beta$ & $<r^2>^{1/2}$ & RMS-exp \\
\colrule
$^{9}$Be         & 2.08 & 0.375  & 1.183 & 2.540  & 2.52(1)  \\
$^{144}$Sm       & 5.829& 0.54   & 0     & 4.941  & 4.947(9) \\
$^{208}$Pb($\pi$)& 6.60 & 0.546  & 0     & 5.500  & 5.503(2) \\
$^{208}$Pb($\nu$)& 6.82 & 0.546  & 0     & 5.66   & 5.66    \\  
\colrule
\end{tabular}
\end{table}

\subsection{Multipole expansion of density}

The density of the deformed nucleus is expanded on Legendre
polynomials,
\beq
\rho(r,\theta') = \sum_\lambda \rho_\lambda(r) \ 
P_\lambda(\cos(\theta')),
\label{rhopl}
\eeq
where $\rho_\lambda(r)$ is calculated numerically,
\beq
\rho_\lambda(r) = \frac{2\lambda+1}{2} \
\int_{-1}^1 dx \ P_\lambda(x) \rho({\bf r}).
\eeq
The multipole expansion of the Fourier transform of the 
density is
\beq
\rho({\bf k}) = \sum_\lambda i^{-\lambda} \rho_\lambda(k) \ 
P_\lambda(\cos(\theta_k'),
\label{rhok}
\eeq
where $\theta_k'$ is the angle between ${\bf k}$ and 
the direction ${\bf e}$ of the symmetry axis,
and
\beq
\rho_\lambda(k) = 4\pi \int_0^\infty dr \ r^2 \
\rho_\lambda(r) \ j_\lambda(kr).
\eeq

The above expressions are used in the next section to calculate 
the double-folding potential. They are also used to calculate the 
electric multipole moments of the deformed charge density 
$\rho_c(r,\theta')$,
\beq
M(E\lambda\mu) = M(E\lambda) \ Y_{\lambda\mu}({\bf e}),
\eeq
where 
\beq
M(E\lambda) = 
\frac{4\pi}{2\lambda+1} \ 
\ \int_0^\infty dr \ r^{\lambda+2} \ \rho_{c\lambda}(r).
\eeq
The intrinsic quadrupole moment $Q_0$ is traditionally
defined as 2$M(E2)$. 

\begin{figure}
\includegraphics[width = 7cm]{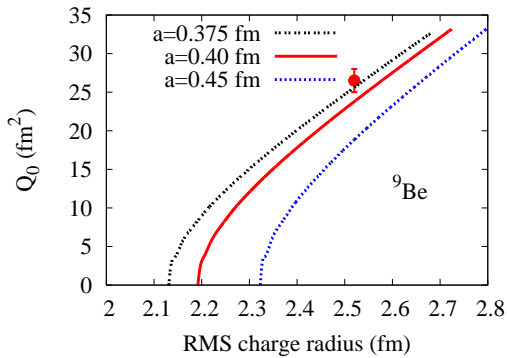}
\caption{\label{q2rmsr} (Color online)
Correlation between the intrinsic quadrupole moment 
and RMS charge radius of $^9$Be. The curves were obtained by varying 
$\beta_2$ for the fixed radius parameter $R$ = 2.08 and the three 
values of the diffuseness indicated in the figure.}
\end{figure}

\begin{figure}
\includegraphics[width = 8cm]{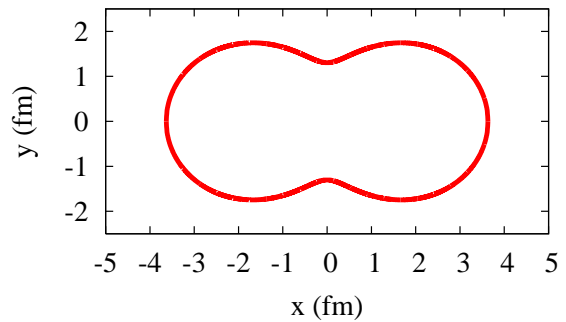}
\caption{\label{shape} (Color online)
The shape of $^9$Be, Eq. (\ref{shaperad}), derived 
by calibrating the density to reproduce the measured intrinsic
quadrupole moment $Q_0$ and RMS charge radius.}
\end{figure}

\subsection{Calibration of the density of $^9$Be} 

The parameters of the density of the deformed $^9$Be nucleus were 
adjusted so that both the intrinsic quadrupole moment and the RMS 
charge radius agree with experiments. That was achieved as follows.
The mean square charge radius of $^9$Be, 
\beq
\langle r^2 \rangle = \frac{4\pi}{Z} \
\int_0^\infty dr \ r^4 \ \rho_{c,\lambda=0}(r),
\eeq
and the intrinsic quadrupole moment, 
\beq
Q_0 = 2M(E2) =
\frac{4\pi}{5} \ 
\ \int_0^\infty dr \ r^{4} \ \rho_{c,\lambda=2}(r).
\eeq
were calculated as functions of the deformation parameter $\beta_2$
for a fixed radius $R$ = 2.08 fm and for three values of the diffuseness.
The results are shown in Fig. \ref{q2rmsr} as a correlation between the
RMS charge radius and $Q_0$. It is seen that the curve which is based on 
the diffuseness $a$ = 0.375 fm passes trough the experimental values, and 
agreement with both values is achieved for $\beta_2$ = 1.183. 
This is the value that will be used in the following, and the shape it
produces according to Eq. \ref{shaperad} looks almost like two touching
$\alpha$ particles as illustrated in Fig. \ref{shape}. In fact, the
intrinsic quadrupole moment of $^9$Be, $Q_0$ = 26.45(20) fm$^2$, is
almost identical to the calculated quadrupole moment of the unbound 
nucleus $^8$Be. The published value obtained in 
Variational Monte Carlo (VMC) calculations is
26.6(3) fm$^2$ \cite{wiringa1}.
The intrinsic quadrupole moment of $^{10}$Be, on the other hand, is
slightly smaller; the value one obtains from 
the measured $B(E2)$ value of the lowest 2$^+$ excitation \cite{ENDSF}
is $Q_0$ = 22.9 fm$^2$.  

\section{Double-folding potential}

Having adopted the rotor model for $^9$Be and determined the densities
of the projectile and the spherical $^{144}$Sm and $^{208}$Pb targets, 
one can now use the double-folding technique to calculate the potential
that will be used in the coupled-channels calculations.
The double-folding potential is defined by
\beq
U({\bf r}) = \int d{\bf r}_1\int d{\bf r}_2 \
\rho({\bf r}_1,{\bf e}) \ \rho_T({\bf r}_2) \
v(|{\bf r}_2+{\bf r}-{\bf r}_1|),
\eeq
where $v$ is the effective nucleon-nucleon interaction and ${\bf r}$ 
is the relative distance between projectile and target.
The target density $\rho_T$ is assumed to be spherical whereas
the density of the projectile $^9$Be is deformed and parametrized
as described in the previous section. 
 
The double-folding potential is calculated most conveniently from 
the Fourier transforms of the densities according to the expression 
\cite{misi1},
\beq
U({\bf r}) = \int \frac{d{\bf k}}{(2\pi)^3} \ 
\rho({\bf k}) \ \rho_T({\bf -k}) \
v(k) \ e^{i{\bf kr}}.
\label{unk}
\eeq
Inserting the expression (\ref{rhok}) for the deformed projectile 
and spherical target densities into Eq. (\ref{unk}) one obtains
\beq
U({\bf r}) = U(r,\theta') =
\sum_\lambda 
U_{\lambda}(r) \ P_\lambda(\cos(\theta')),
\label{ulam}
\eeq
where 
$\theta'$ is the angle between {\bf r} and {\bf e} and
\beq
U_{\lambda}(r) = \frac{1}{2\pi^2} \
\int dk \ k^2 \rho_\lambda(k) \ \rho_T(k) \
v(k) \ j_\lambda(kr).
\eeq

The double-folding calculation of the ion-ion potential and its
multipole expansion, Eq. (\ref{ulam}), will be based on the M3Y effective 
interaction, supplemented with a repulsive term that simulates the 
effect of nuclear in-compressibility. This method has been applied 
previously by Mi\c sicu and Greiner \cite{misi2} to calculate the
fusion between spherical and deformed nuclei. It was also 
used in Ref. \cite{misih} to explain the hindrance in the fusion
of spherical nuclei at extreme subbarrier energies.

The repulsive term in the effective $NN$ interaction is parametrized
as a contact interaction,
\beq
v_{NN}^{(\rm rep)}({\bf r}) = v_r \delta({\bf r}),
\label{repul}
\eeq
and the densities that are used in the associated double-folding
calculation have the same radius as shown in Table I but the 
diffuseness $a_r$ is usually chosen much smaller \cite{abe}.
The value chosen here is $a_r$ = 0.3 fm.

The Coulomb interaction interaction can also be generated from 
Eq. (\ref{unk}) simply by replacing $v(k)$ by the Fourier
transform $4\pi e^2/k^2$ of the proton-proton Coulomb interaction,
and replacing the nuclear densities with the charge densities 
$\rho_c$ of projectile and target. 
The results has the same form as Eq. (\ref{ulam}),
\beq
U_C(r,\theta') = 
\sum_\lambda U_{\lambda}^C(r) \ P_\lambda(\cos(\theta')), 
\eeq
where
\beq
U_{\lambda}^C(r) = \frac{1}{2\pi^2} \
\int dk \ k^2 \rho_{c,\lambda}(k) \ \rho_{c,T}(k) \
\frac{4\pi e^2}{k^2} \ j_\lambda(kr).
\eeq
For large separations of projectile and target this interaction 
approaches the usual monopole-multipole interaction,
\beq
U_{\lambda}^C(r) =
\frac{Z_Te^2M(E\lambda)}{r^{\lambda+1}}.
\eeq

\subsection{Matrix elements}

Having expressed the total interaction $U(r,\theta')$ (Coulomb + Nuclear)
in terms of the multipole expansion Eq. (\ref{ulam}), one can now easily
calculate the diagonal as well as off-diagonal couplings between states 
in the ground state rotational band of $^9$Be. All one needs to calculate 
is matrix elements of the Legendre polynomials,
$$P_\lambda(cos(\theta')) =
\sum_\mu
D_{\mu 0}^{\lambda *}({\hat r}) \
D_{\mu 0}^\lambda ({\bf e}).
$$ 
The matrix elements between different states are
$$
\langle KI'M'| 
P_\lambda(cos(\theta'))
|KIM\rangle = 
$$
\beq
\sqrt{\frac{2I+1}{2I'+1}} \
\sum_\mu D^{\lambda*}_{\mu 0}({\hat r}) \
\langle IM \ \lambda \mu| I'M'\rangle \
\langle IK \ \lambda 0| I'K\rangle.
\label{plmat}
\eeq
\begin{figure}
\includegraphics[width = 8cm]{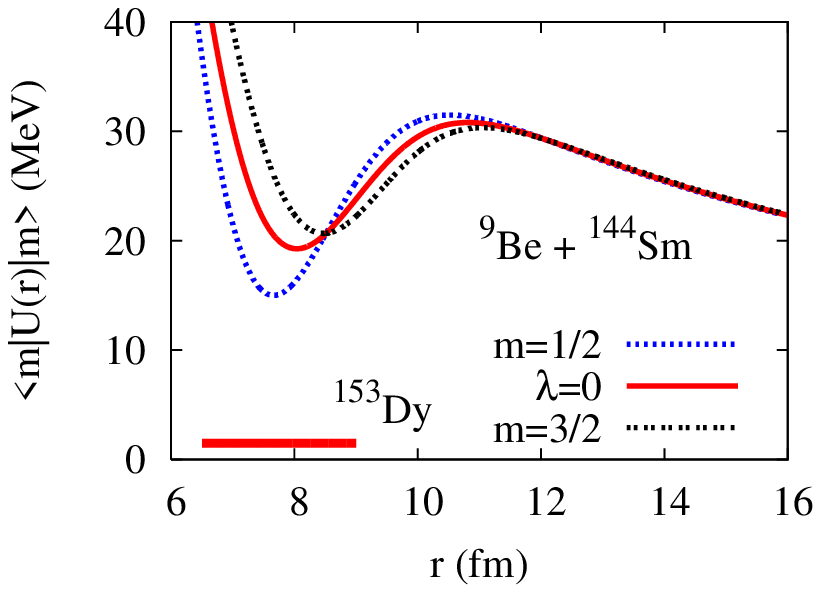}
\includegraphics[width = 8cm]{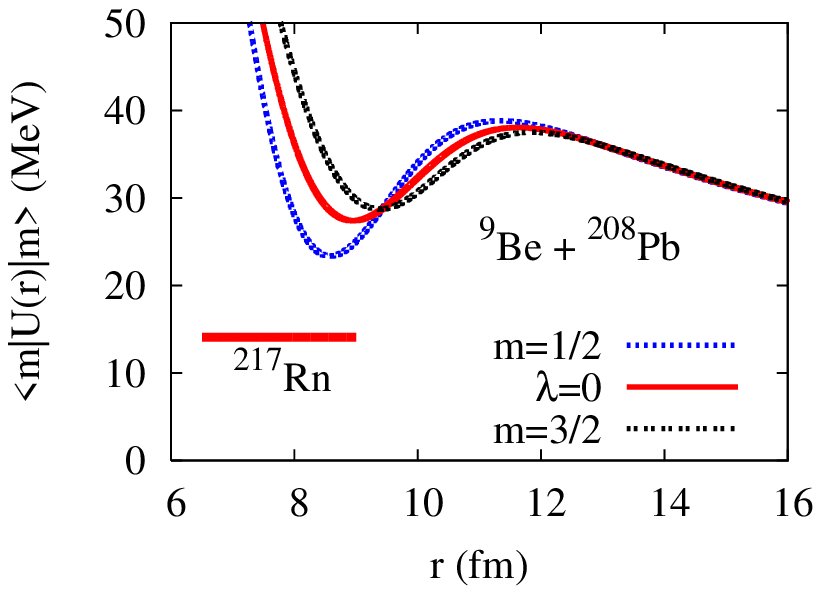}
\caption{\label{spupot} (Color online)
Entrance channel potentials for 
$^9$Be+$^{144}$Sm and $^9$Be+$^{208}$Pb, respectively,
obtained for the repulsive strength $v_r$ = 410 MeV fm$^3$.  
The solid curve is the monopole potential; the dashed curves 
are the entrance channel potentials for the magnetic quantum 
numbers $m$ = 1/2 and 3/2. The ground state energies of the two 
compound nuclei, $^{153}$Dy and $^{217}$Rn, are indicated.} 
\end{figure}
The calculation is even simpler in the rotating frame approximation,
which is used in the coupled-channels calculations described in the 
next section. In this approximation one assumes that ${\bf r}$ 
(the relative distance between projectile and target) defines 
the $z$-axis. 
The angle $\theta'$ is then identical to the angle $\theta_e$ of the
symmetry axis with respect to the $z$-axis. Since
$D_{\mu 0}^{\lambda}({\hat z})$ = $\delta_{\mu 0}$ this implies 
that $\mu$=0 is the only non-zero term in Eq. (\ref{plmat}).

The total potentials one obtains for the 
two systems $^9$Be+$^{144}$Sm and $^9$Be+$^{208}$Pb 
are shown in Fig. \ref{spupot}. Shown are for each system the 
monopole potential (solid line) and the entrance channel potentials
for the magnetic quantum numbers $m$=1/2 and 3/2, of the $3/2^-$, 
$K$=3/2 ground state of $^9$Be.
All three potentials were obtained with the strength $v_r$ = 410 
MeV fm$^3$ of the repulsive effective $NN$ interaction 
(which is determined in subsection V.A.)

The magnetic quantum numbers $m$=1/2 and 3/2 used in Fig. \ref{spupot}
refer to a $z$-axis that points in the direction of the relative 
position of projectile and target. 
The $m$=3/2 channel therefore corresponds to an orientation where 
the tip of the deformed $^9$Be points towards the target, whereas
the $m$=1/2 corresponds to the belly pointing towards the target.
Consequently, the Coulomb barrier for the $m$=3/2 entrance channel
is lower than the barrier for $m$=1/2.
Another observation is that the 
potential pocket is deeper for $m$=1/2 than for $m$=3/2.
This is a consequence of a larger radius of curvature and a stronger 
nuclear attraction for the $m$=1/2 belly configuration.  


\section{Model of $^9$Be induced fusion}

The cross sections for the complete fusion of $^9$Be with a heavy
target are calculated in the coupled-channels approach.
The complete fusion is simulated by ingoing wave boundary conditions 
that are imposed in all channels at the minimum of the pocket in
the entrance channel potential. 
The coupled equations are solved in the rotating frame approximation
\cite{Ron,Tamu,Land}, where the $z$-axis points in the direction of the 
separation vector ${\bf r}$ between the reacting nuclei. 
This approximation is also called the isocentrifugal approximation
\cite{Ron} because the centrifugal potential is the same in all channels
and equal to the centrifugal potential in the entrance channel.
The total magnetic quantum number $m$ is also preserved in this 
approximation (see Ref. \cite{Esb03}). 
Since the ground state of the target is a $0^+$ state and the ground 
state of $^9$Be is a $3/2^-$ state, one would have 
to solve the coupled equations four times, for $m$ = $\pm 1/2$ and 
$\pm 3/2$. For symmetry reasons it is actually sufficient to calculate
the fusion cross section twice, for $m$ = 1/2 and 3/2, and the average
fusion cross section will be compared with data.

The effect of the decay of the excited state in $^9$Be is included
explicitly in the coupled equations by employing the 
complex excitation energies $E_I-\frac{i}{2}\Gamma_I$. 
The decay widths $\Gamma_I$ are shown in the 3rd column of Table I;
it is seen that the width of the 7/2$^-$ state is very large. 
The dominant decay mode of the $7/2^-$ state is neutron emission,
and 55\% of it populates the $2^+$ excited state of $^8$Be \cite{ENDSF}. 
The latter state has an excitation energy  of 3.03 MeV and a large 
width of 1.51 MeV, with an exclusive decay into two $\alpha$ 
particles \cite{ENDSF}. The two $\alpha$ particles are emitted 
back to back in the $^8$Be rest frame, so if one of them is emitted 
towards the target, the other $\alpha$ partner will recoil away from 
the target nucleus and will most likely escape. 
It is therefore assumed that the decay of the $7/2^-$ 
state will not lead to complete fusion (CF), and the CF will
be calculated as described above from the ingoing flux, whereas
the decay is registered in the absorption cross section.

\begin{figure}
\includegraphics[width = 8cm]{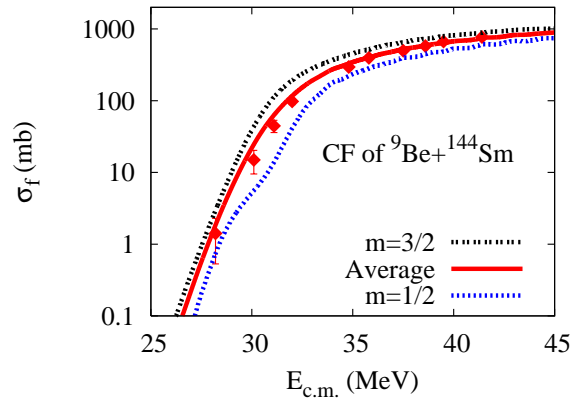}
\caption{\label{besmcfmdp} (Color online)
Measured cross sections for the complete
fusion of $^9$Be+$^{144}$Sm \cite{anu06sm} 
are compared to calculations for $m$ = 1/2 and 3/2, and the 
average cross section (solid curve).
The strength of the repulsive interaction
was set to $v_r$ = 410 MeV fm$^3$.}
\end{figure}

The data for the complete fusion of $^9$Be+$^{144}$Sm \cite{anu06sm}
are compared in Fig. \ref{besmcfmdp} to the results of coupled-channels 
calculations. The calculated CF cross sections were derived from the
ingoing flux as described above.
The top dashed curve is the cross section for the $m$ = 3/2 magnetic substate, 
the lower dotted curve is for $m$ = 1/2. It is seen that the curve for 
$m$ = 3/2 dominates the CF at all energies, consistent with the lower 
Coulomb barrier for this magnetic substate (see Fig. \ref{spupot}.)  
The data should be compared to the solid curve which is the average 
of the CF cross sections for $m$ = 1/2 and 3/2. The comparison is
discussed in more detail in subsection V.A.

\subsection{Incomplete fusion}

The decay of the $7/2^-$ state will end up in the breakup of $^9$Be.
The precise outcome of the decay in terms of 
incomplete fusion or breakup is not so clear. It would require a
multi-cluster description to follow the two $\alpha$ particles after
the decay. As mentioned earlier, it is unlikely that both $\alpha$ 
particles fuse with the target because they are emitted back to back. 
However, it is possible that one of them will fuse with the target 
nucleus and lead to incomplete fusion (ICF).
It should also be emphasized that there are other sources of ICF, 
for example, the neutron transfer from $^9$Be, which are not included 
in the coupled-channels calculations presented here.

The calculated cross section for the decay of the excited states
will in the following be referred to as the absorption cross section. 
In view of the above discussion one should not expect that the 
absorption would account for the measured ICF cross sections 
but it is clearly of interest to compare the two cross sections. 
The experimental total fusion (TF) cross section is the sum of the 
CF and ICF cross sections. It will be compared to the calculated 
TF cross section, which is the sum of the CF and the absorption 
cross sections.


\section{Comparison to measurements}

\begin{figure}
\includegraphics[width = 7cm]{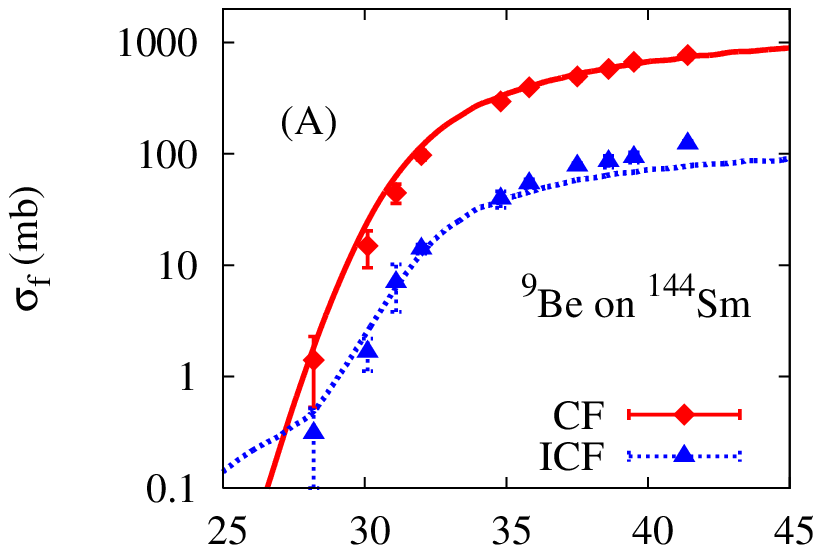}
\includegraphics[width = 7cm]{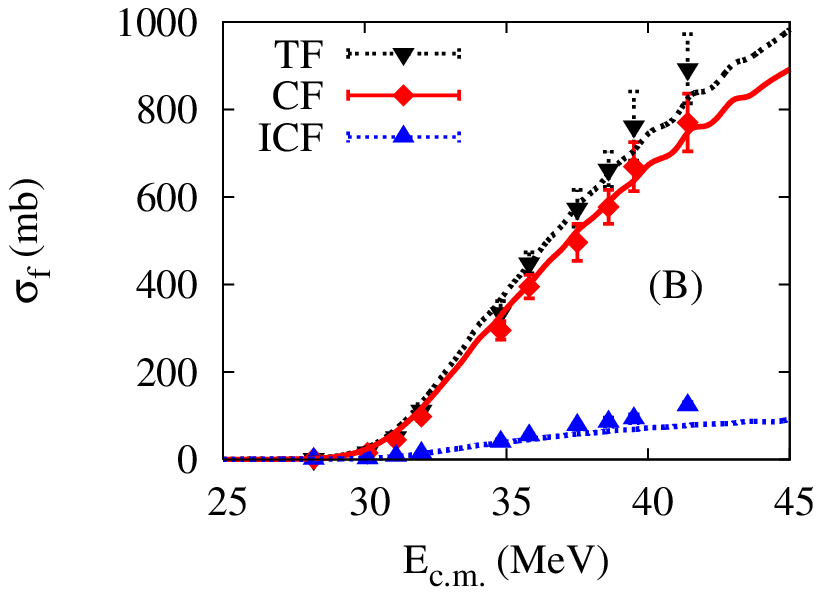}
\caption{\label{besmf} (Color online) 
Measured complete (CF) and incomplete (ICF)
fusion cross sections for $^9$Be+$^{144}$Sm \cite{anu06sm} are 
compared in (A) to the calculated cross sections for CF (solid) and 
absorption (dashed curve). The linear plot in (B) also shows the total 
fusion (TF) cross section.}
\end{figure}
\begin{figure}
\includegraphics[width = 8cm]{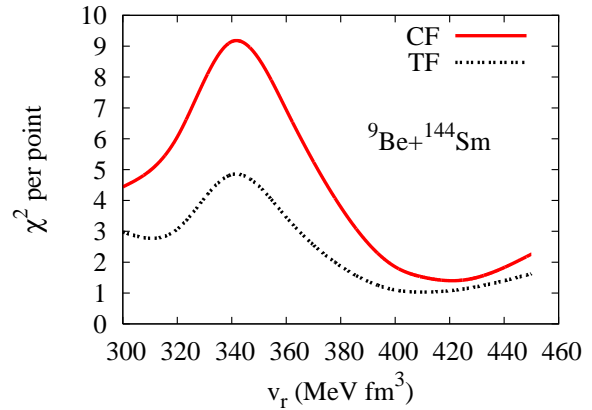}
\caption{\label{xki2} (Color online)
The $\chi^2$ per data point for the complete (CF)
and total fusion (TF) of $^9$Be+$^{144}$Sm are shown as functions of 
the strength $v_r$ of the repulsive interaction, Eq. (\ref{repul}).}
\end{figure}

The results of coupled-channels calculations that are based on 
the model presented in the previous sections are compared in the 
following to the data for the complete (CF) and incomplete fusion 
(ICF) of $^9$Be+$^{208}$Pb \cite{anu99cf,anu04pb} and 
$^9$Be+$^{144}$Sm \cite{anu06sm}.
Both targets are spherical, closed shell nuclei and the excitation 
of these nuclei is relatively weak compared to the excitation of
$^9$Be and they will be ignored.
The fusion with the $^{144}$Sm target is discussed first
because the couplings are weaker in this case and the adopted 
model is therefore expected to be more successful. This case 
will also provides the opportunity to calibrate the repulsive part
of the effective $NN$ interaction.

\subsection{Fusion of $^9$Be+$^{144}$Sm.}

The calculated cross sections for the fusion of $^9$Be 
with $^{144}$Sm are compared in Fig. \ref{besmf}A to the data 
of Ref. \cite{anu06sm}.  The measured and calculated cross 
sections for complete fusion (CF) are seen to be in good
agreement. This was achieved by adjusting the repulsive part of 
the effective $NN$ interaction which is used in the calculation
of the double-folding nuclear potential. The best fit to the data 
is obtained for the strength $v_{r}\approx$ 410 MeV fm$^3$, and 
that is the value that will be used in the following.
The $\chi^2$ per data point is shown in Fig. \ref{xki2} as 
a function of the strength of the repulsive interaction.
There is another solution with a small $\chi^2$ for $v_{r}\leq$ 
300 MeV fm$^3$ but it is unphysical because it produces a pocket 
for $^9$Be+$^{208}$Pb that is deeper than the energy
of the compound nucleus.

The dashed curve in Fig. \ref{besmf}A shows the calculated 
cross sections for the decay of the excited states of $^9$Be;
it is in surprisingly good agreement with the ICF data. 
The good agreement may be accidental but it could also 
indicate that the decay of the excited states of $^9$Be is 
the main source of ICF for the $^{144}$Sm target. 
The best fit to the data for total fusion (TF) is also achieved 
for $v_{r}\approx$ 410 MeV fm$^3$.  This consistency of the 
CF and TF is illustrated in Fig. \ref{xki2} in terms 
of the $\chi^2$ per data point.

The measured and calculated fusion cross sections are compared in
a linear plot in Fig. \ref{besmf}B. It is seen that the different 
components of the measured and calculated fusion cross sections 
are in good agreement.  In particular, the CF is suppressed 
by about 10\% compared to the total fusion (TF) at high energies,
both experimentally and in the calculations. 
This suppression is caused in the coupled-channels calculations 
by the decay of the excited states of $^9$Be. Without any decay
in the coupled-channels calculations, the fusion cross section 
obtained from the ingoing-wave boundary conditions would be  
close to the measured TF cross section. In other words, the 
suppression of the CF compared to the TF requires some sort of 
absorption mechanism, and the decay mechanism suggested here
seems to provide a natural explanation. Let us now investigate 
whether this mechanism can explain the data for the Pb target.



\subsection{Fusion of $^9$Be+$^{208}$Pb.}

\begin{figure}
\includegraphics[width = 7cm]{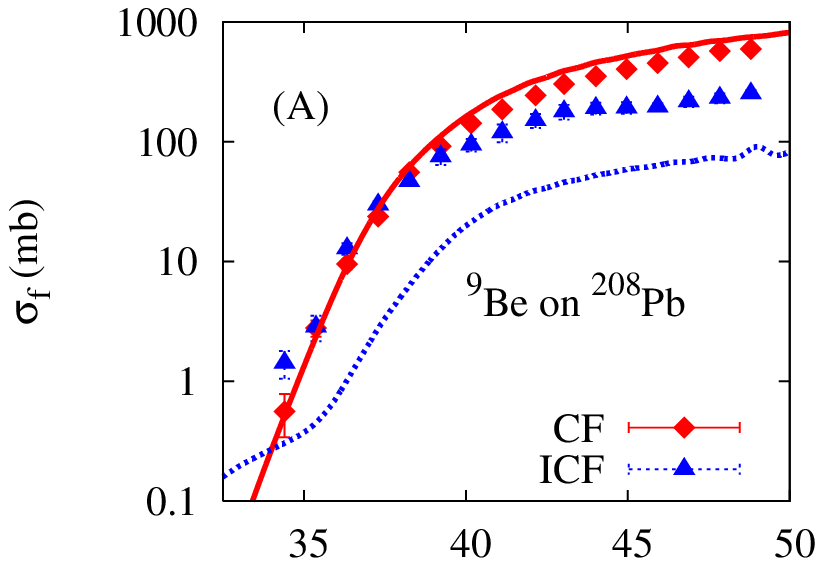}
\includegraphics[width = 7cm]{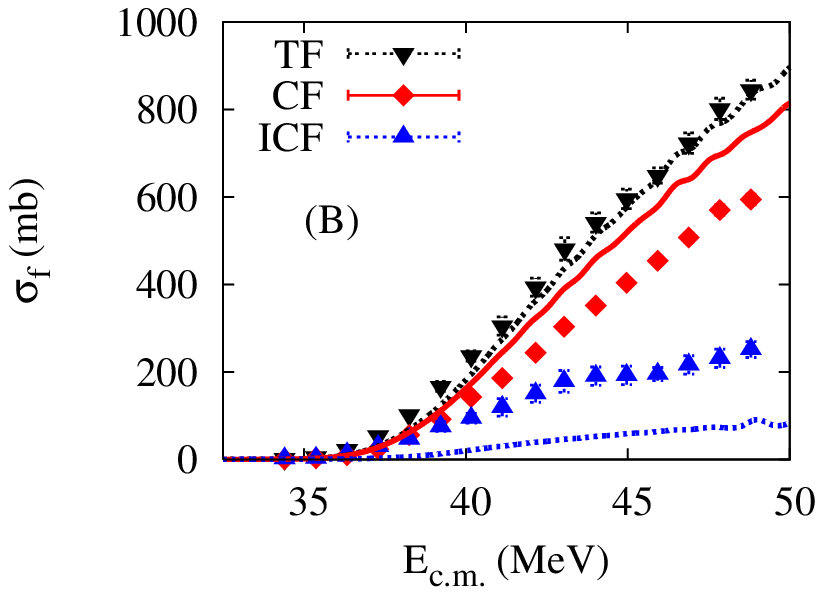}
\caption{\label{bepbf} (Color online)
Measured cross sections for the complete (CF)
and incomplete fusion (ICF) of $^9$Be+$^{208}$Pb \cite{anu04pb} are 
compared to the calculated cross sections for CF (solid curve) and 
absorption (dashed curve). 
The linear plot in (B) also shows the total fusion (TF) cross sections.}
\end{figure}

The results of the coupled-channels calculations of the fusion 
of $^9$Be+$^{208}$Pb are compared in Fig. \ref{bepbf}
to the data of Ref. \cite{anu04pb}. The calculations are
similar to those presented above for $^9$Be+$^{144}$Sm.
It is seen in Fig. \ref{bepbf}B that the calculated  absorption
cross section (due to the decay of excited states) can only 
explain 1/3 of the measured ICF cross section at high energies.
The suppression of the CF compared to the TF cross section is 
about 30\% in the experiment \cite{anu04pb}, whereas the 
calculations only show a 10\% suppression. 
There are evidently other sources of ICF in collisions of $^9$Be
with a $^{208}$Pb target, besides the decay of excited states
considered here.

A 30\% suppression of the CF data was observed in Ref. 
\cite{anu04pb} by comparing to coupled-channels calculations.
The calculations were based on a deformed Woods-Saxon potential 
but did not consider the effects of incomplete fusion nor 
the decay of excited states.  It was shown that a scaling of
the calculated fusion cross section by a factor of 0.7 leads 
to a very good agreement with the CF data at all energies. 
In Fig. \ref{bepbf} it is sufficient to multiply the CF calculation 
by a factor 0.8 in order to match the CF data at high energies.
The reason for the smaller scaling factor is that the decay of 
the excited states has already taken care of a 10\% reduction.

\begin{figure}
\includegraphics[width = 7cm]{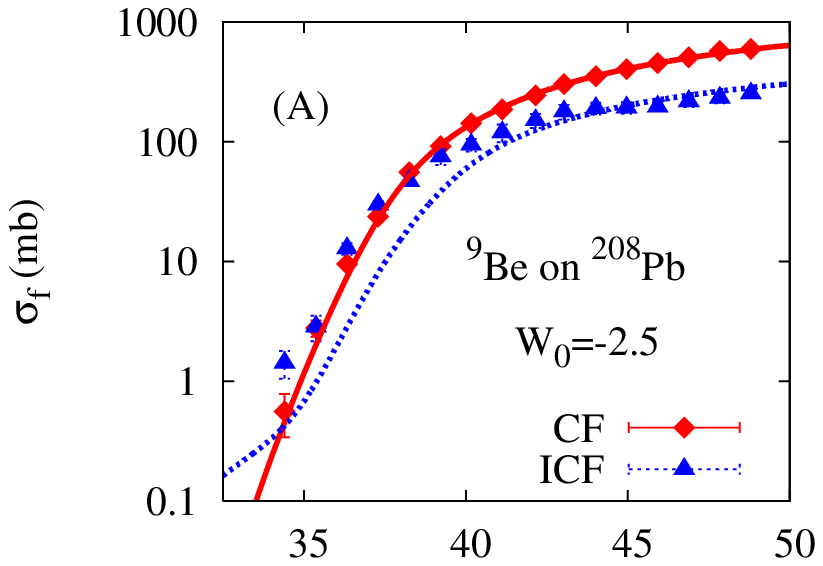}
\includegraphics[width = 7cm]{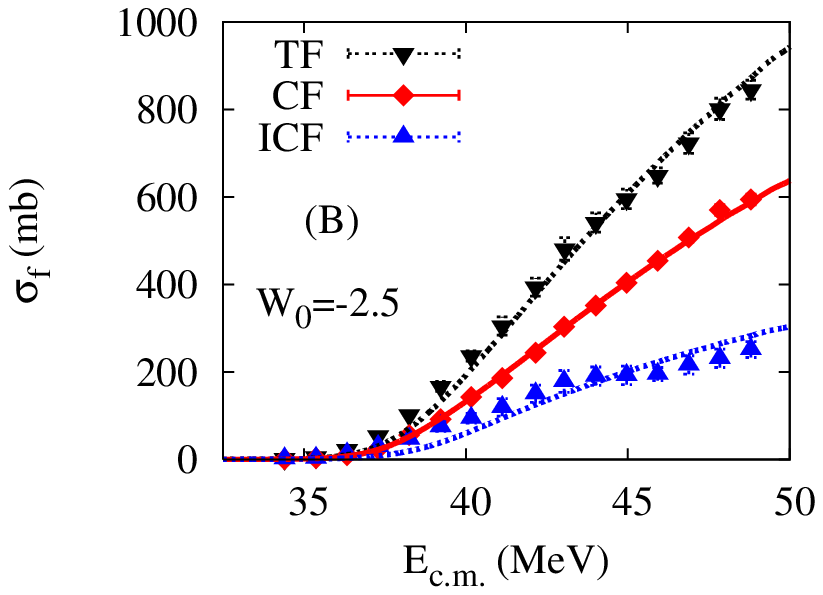}
\caption{\label{bepbfw} (Color online)
Similar to Fig. \ref{bepbf}. 
The calculations include, in addition to the decay of the excited 
states of $^9$Be, a weak imaginary potential with the strength 
$W_0$ = -2.5 MeV.}
\end{figure}

It is often necessary to employ a weak, short-ranged imaginary
potential in order to be able to reproduce the fusion data of 
stable nuclei at high energies by coupled-channels calculations. 
This is particularly the case when the calculations are based on 
a shallow entrance channel potential \cite{Esb07}. Since the 
potentials shown in Fig. \ref{spupot} are relatively
shallow it is of interest to see what is the effect of a weak 
imaginary potential on the fusion of $^9$Be+$^{208}$Pb. 
Let us therefore choose the potential
\beq
W(r) = W_0 \bigl[1+exp((r-R_w)/a_w)\bigr]^{-1}.
\label{wopt}
\eeq
with $a_w$ = 0.2 fm and $R_w$ = 9.5 fm so that it acts near the minimum
of the potential pockets shown in Fig. \ref{spupot}.  The strength $W_0$ 
was adjusted to optimize the fit to the CF data. The best fit is shown 
in Fig. \ref{bepbfw}A; it was achieved for $W_0$ = -2.5 MeV and has a 
$\chi^2/N$=1.4. 

The absorption cross section shown in Fig. \ref{bepbfw}, due to the 
combined effect of the imaginary potential, Eq. (\ref{wopt}), and the 
decay of the excited states, is in good agreement with the ICF data 
at high energies but the discrepancy is large at low energies.
The discrepancy indicates that 
the breakup leading to ICF must take place at larger separations of 
projectile and target than assumed in the potential, Eq. (\ref{wopt}).
It may be possible to construct a more realistic imaginary potential 
(of the volume plus surface type) but that idea will not be pursued here.

There are other reaction channels that could be a source of ICF.
Examples are the dissociation of $^9$Be induced by the neutron 
emission into the continuum and into bound states of the target 
nucleus. Both processes produce $^8$Be which 
decays into two $\alpha$-particles, and one of them could end up as 
incomplete fusion. These reaction mechanisms should be studied 
theoretically in detail in order to develop a better understanding 
and description of the breakup, complete and incomplete fusion.
In fact, a recent experiments \cite{rafiei} show that the $^9$Be 
breakup following neutron transfer dominates the total breakup yield. 


\section{Conclusions}

The complete fusion of $^9$Be with spherical target nuclei was calculated
in the coupled-channels approach. It was shown that the $B(E2)$ values for 
the excitation of the ground state rotational band of the $^9$Be nucleus
can be described quite well in the rotor model. This feature was exploited 
in the calculation of matrix elements of the interaction between the 
deformed projectile and a spherical target. 

The interaction of the deformed $^9$Be projectile with a spherical target 
was calculated using the double-folding technique and an effective M3Y 
$NN$ interaction, which was supplemented with an adjustable repulsive term. 
The deformed density of $^9$Be was determined so the measured quadrupole 
moment and RMS charge radius were reproduced.
The densities of the spherical targets were calibrated to reproduce 
the measured charge radii; the radius of the neutron density in $^{208}$Pb
was calibrated to be consistent with the neutron skin thickness predicted 
by Skyrme Hartree-Fock calculations and with the value extracted from 
measurements of antiprotonic $^{208}$Pb atoms.

The double-folding potential was applied in coupled-channels 
calculations of the fusion of $^9$Be with $^{144}$Sm. 
The decay of the excited states of $^9$Be was included explicitly 
in terms of complex excitation energies, whereas excitations of
the target were ignored for simplicity.
The repulsive part of the effective $NN$ interaction, which essentially 
is the only parameter that remains to be determined, was adjusted 
to produce an optimum fit to the complete fusion data. 
The calculated cross sections for the decay of the excited states in 
$^9$Be turned out to be in very good agreement with the incomplete 
fusion data. 

Having determined all of the parameters of the theory, coupled-channels
calculations were performed for the fusion of $^9$Be+$^{208}$Pb.
The complete fusion data were suppressed at high energies by 20\%
compared to the predicted cross sections, and the incomplete fusion data 
were a factor of three larger than the calculated cross section for the 
decay of the excited states.  
There are obviously other reaction mechanisms, besides the 
decay of excited states, that are responsible for the large incomplete 
fusion cross sections that have been measured for the lead target.
A likely candidate is the neutron transfer from $^9$Be to bound states
in the target and to continuum states. This is also the conclusion of
a recent experimental investigation by Rafiei et al. \cite{rafiei}.

{\bf Acknowledgments}
This work was supported by the U.S. Department of Energy,
Office of Nuclear Physics, under Contract No. DE-AC02-06CH11357.

\end{document}